\documentclass[12pt]{iopart}
\usepackage{graphicx}

\begin{document}

\title[$J/\psi$ hadron interaction in vacuum
and in QGP]{$J/\psi$ hadron interaction in vacuum and in QGP}

\author{Su Houng Lee, Yongjae Park, Kyung-Il Kim and Taesoo Song}

\address{Institute of Physics and Applied Physics,
Yonsei University, Seoul 120-749, Korea}
\ead{suhoung@yonsei.ac.kr}
\begin{abstract}
Motivated by the recent lattice data that $J/\psi$ will survive up
to 1.6$T_c$, we calculate the thermal width of $J/\psi$ at finite
temperature in perturbative QCD.  The inputs of the calculation
are the parton quarkonium dissociation cross sections at the NLO
in QCD, which were previously obtained by Song and Lee, and a
gaussian charmonium wave function, whose size were fitted to an
estimate by Wong by solving the schrodinger equation for
charmonium in a potential extracted from the lattice at finite
temperature. We find that the total thermal width above 1.4$T_c$
becomes larger than 100 to 200 MeV, depending on the effective
thermal masses of the quark and gluon, which we take it to vary
from 600 to 400 MeV.
\end{abstract}


\section{Introduction}
The original claim by Matsui and Satz\cite{Matsui86}, that
$J/\psi$ suppression is a signature of quark gluon plasma, has
witnessed a number of landmark developments, every aspect of which
has to be taken into account consistently in confronting the
recent RHIC data\cite{Phenix1}, and in predicting results for LHC.
Among these theoretical developments are the phenomenologically
successful statistical model for $J/\psi$
production\cite{Gorenstein99,PBM99,PBM06}, based on a coalescence
assumption near $T_c$\cite{PBM06}, and the recent lattice
calculations, showing strong evidence that the $J/\psi$ will
survive up to 1.6$T_c$\cite{Hatsuda03,Hatsuda04,Datta03,Datta05}.
While these two results seem at odd with each other, it only
suggests that one still needs a more detailed understanding of the
properties of heavy quark system in the quark gluon plasma,
especially for temperatures between $T_c$ and $1.6T_c$, before a
consistent picture of $J/\psi$ suppression in heavy ion collision
is achieved.

In this respect, an  important quantity to investigate is the
effective thermal width, and/or the effective dissociation cross
section of $J/\psi$ in the quark gluon plasma.  Except for its
existence, the present lattice results are far from making
definite statements on the thermal width for Charmonium states
above $T_c$.   Hence, in this work, we will use the perturbative
QCD approach to calculate the thermal width. So far, such
calculations have been limited to dissociation processes by gluons
to LO\cite{KS94,AGGA01,Wong04}, because the elementary
$J/\psi$-parton dissociation cross section was available only to
that order\cite{Peskin79,BP79}.   Recently, two of us have
performed the dissociation cross section calculation to NLO in
QCD\cite{SL05}.    Here, we will implement the  NLO formula, to
calculate the corresponding thermal width of Charmonium states in
the quark gluon plasma.

The NLO calculation of $J/\psi$-parton dissociation calculations
involves collinear divergence.  When applying this elementary
cross section to dissociation by hadrons, the collinear divergence
is cured by mass factorization, which renormalizes the divergent
part of the cross section into the parton distribution function of
the hadron.   Such complications disappear at finite temperature,
as the thermal masses of the partons automatically renders the
divergence finite.

\section{Quarkonium hadron interaction in QCD}

Let us begin with some introduction on the propagation of heavy
quarks in the QCD vacuum. The propagation of a heavy quark can be
approximated by a perturbative quark propagation with a
perturbative gluon insertion, which probes the non perturbative
gluon field configuration in the QCD vacuum.   Hence, the full
heavy quark propagator is,
\begin{eqnarray}
iS^A(q)=iS(q)+iS(q)(-ig A  \!\!\!/ )iS(q) \cdot \cdot \cdot ,
\label{propagator}
\end{eqnarray}
where, $iS(q)=i/(q \!\!\!/ -m)$ and $m$ is the heavy quark mass.
The description in Eq.(\ref{propagator}) is valid even for
$q\rightarrow 0$, because $m\gg A \sim \Lambda_{QCD}$, where in
the end only gauge invariant combination of the gauge field $A$
will remain after taking the vacuum expectation value.

\begin{table}[ht]
\caption{\label{table1}Physical processes involving two heavy
quarks. }
\begin{center}
\begin{tabular}{lll}
\br
$q^2$ &  Process & Expansion parameter\\
\mr
0  & Photo production of open charm   & $\frac{ \Lambda_{QCD}^2 }{4m^2}$ \\
$-Q^2 <0$ & QCD sum rules for heavy quark system & $\frac{ \Lambda_{QCD}^2 }{4m^2+Q^2 }$ \\
$m_{J/\psi}^2 >0$ & Dissociation cross section of bound states &
$\frac{ \Lambda_{QCD}^2 }{4m^2-m_{J/\psi}^2 }$ \\
\br
\end{tabular}
\end{center}
\end{table}

The propagation of a system composed of a heavy quark and an
antiquark can also be approximated by combined perturbative heavy
quark propagator with gluon insertions. However, since there a two
heavy quarks involved, based on the operator product expansion,
the propagation can typically be written in the following form.
\begin{eqnarray}
\Pi(q)=....+\int_0^1 dx {F(q^2,x) \over (4m^2-q^2-(2x-1)^2q^2)^n}
\langle G^n \rangle  \cdot \cdot \cdot, \label{twoquark}
\end{eqnarray}
where, $F(q^2,x)$ is a function depending on the structure of the
two quark system and $\langle G^n \rangle \sim \Lambda_{QCD}^{2n}$
denotes the typical gauge invariant expectation value of gluonic
operator of dimension $2n$.  The integration variable $x$ can be
thought of as the momentum fraction carried by one of the heavy
quark. Here, one notes that such perturbative expansion is valid
when $4m^2-q^2 \gg \Lambda_{QCD}^2$.  The  cases where this
condition is satisfied and perturbative QCD treatments are
possible are summarized in Table 1.  In the last line of Table
\ref{table1}, $4m^2-m_{J/\psi}^2 \approx (2m+m_{J/\psi})
\epsilon_0$, where $\epsilon_0$ is the binding energy of the
$J/\psi$.   In QCD if $m\rightarrow \infty$, the bound state
becomes Coulombic and $\epsilon_0=m[N_c g^2/(16 \pi)]^2 \gg
\Lambda_{QCD}$ in the large $N_c$ limit.  Therefore, the expansion
parameter becomes small and the dissociation cross can be
calculated using perturbative QCD.

\section{Thermal width of $J/\psi$}

The LO invariant matrix element for the $J/\psi$ dissociation
cross section by gluon was first obtained by Peskin\cite{Peskin79}
and rederived by one of us\cite{OKL02} using the Bethe-Salpeter
equation.  The NLO invariant matrix element was derived by two of
us\cite{SL05}.  The dissociation by quarks comes in as part of the
NLO calculation.  The thermal width of $J/\psi$ is calculated by
folding the elementary cross section with the thermal parton
distribution function.
\begin{eqnarray}
\Gamma_T=deg \int \frac{d^3k}{(2\pi)^3} n(k_0) v \sigma(s),
\label{width}
\end{eqnarray}
where $deg$ is the degeneracy of the partons, $n(k_0)$ is the
thermal quark or gluon distribution function and $\sigma(s)$ is
the elementary cross section, with $s$ being the square of the sum
of the gluon and charmonium four momentum, and $v$ their relative
velocity.

In both the LO and NLO calculations, $\sigma(s)$ is proportional
to $|\partial \phi(p) /\partial p|^2$, where $\phi$ is the
charmonium wave function and $|p|$ is related to the incoming four
momentum that breaks the $J/\psi$ into $\bar{c} c$. As temperature
increases, $n(k_0)$ favors higher $k_0$ or $|k|$, while the wave
function favors smaller $|p|$, because the wave function becomes
shallow and the momentum space wave function becomes smaller, for
which we take the result given in ref\cite{Wong04}.
\begin{figure}
\begin{center}
\includegraphics[height=5cm,width=8cm]{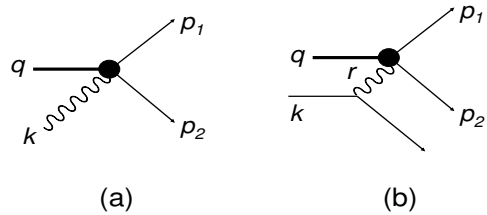}
\caption{\label{fig1} The (a) LO diagram and (b) quark induced NLO
diagram. $q,p_1,p_2,k$ are the four momentum of the
$J/\psi,c,\bar{c}$ and the incoming parton respectively.}
\end{center}
\end{figure}

For the LO calculation shown in Fig 1 (a),
$k_0=\epsilon_0+p^2/(2m_c)$. Hence, larger $k_0$ means larger $p$.
That means at higher temperature, although there are more
energetic gluons, these are inefficient in breaking the charmonium
states.  Moreover, if one introduces thermal gluon mass, the LO
width becomes even smaller.

For the NLO calculation, Fig 1 (b) and the corresponding diagram
where the incoming quark is replaced by a gluon, have collinear
divergences, which in the vacuum is cured by mass
factorization\cite{SL05}.  At finite temperature, the problem can
be avoided naturally by introducing thermal masses, which we take
here to be from 600 MeV\cite{LH97} to 400 MeV, the smaller value
giving an upper limit to the effective width. Moreover, even an
energetic parton can radiate a gluon with small four momentum ($r$
in Fig 1 (b)), which can effectively dissociate the $J/\psi$ into
open charms. Fig 2 shows the effective width of $J/\psi$ obtained
by folding the elementary NLO quark (a) and the corresponding
gluon (b) dissociation cross section obtained in ref\cite{SL05} by
a massive thermal parton distribution.  As can be seen from the
figure, even with a upper (lower) limit of 600 (400) MeV thermal
mass, the sum of thermal width due to quarks and gluons becomes
larger than 100 (200) MeV above 1.4$T_c$, where we take $T_c=170$
MeV.  Hence, although the $J/\psi$ might start forming at 1.6
$T_c$, the effective thermal width is very large and will not
accumulate until the system cools down further\cite{PBM06}.

\begin{figure}[h]
\begin{minipage}{18pc}
\includegraphics[width=16pc]{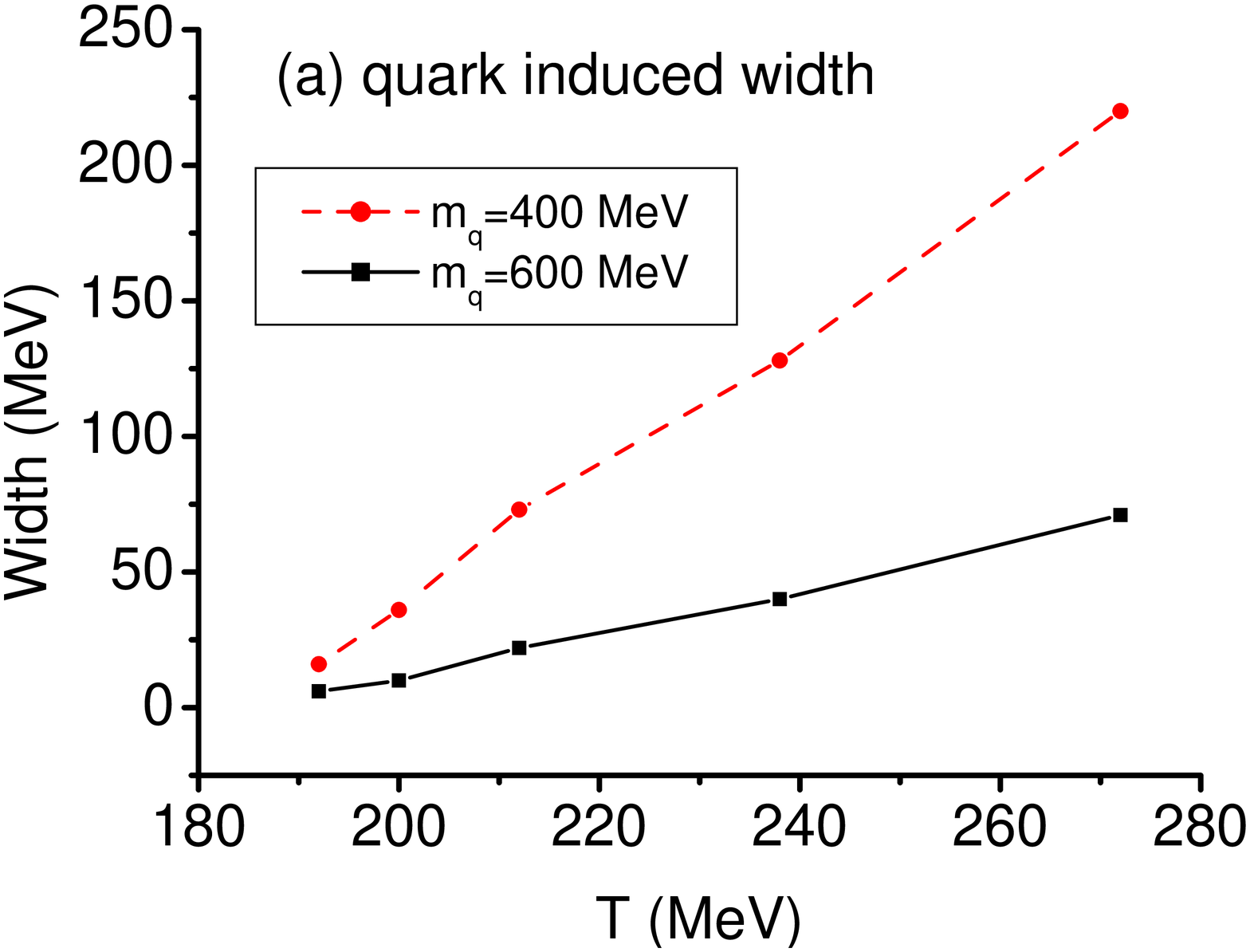}
\end{minipage}\hspace{2pc}%
\begin{minipage}{18pc}
\includegraphics[width=16pc]{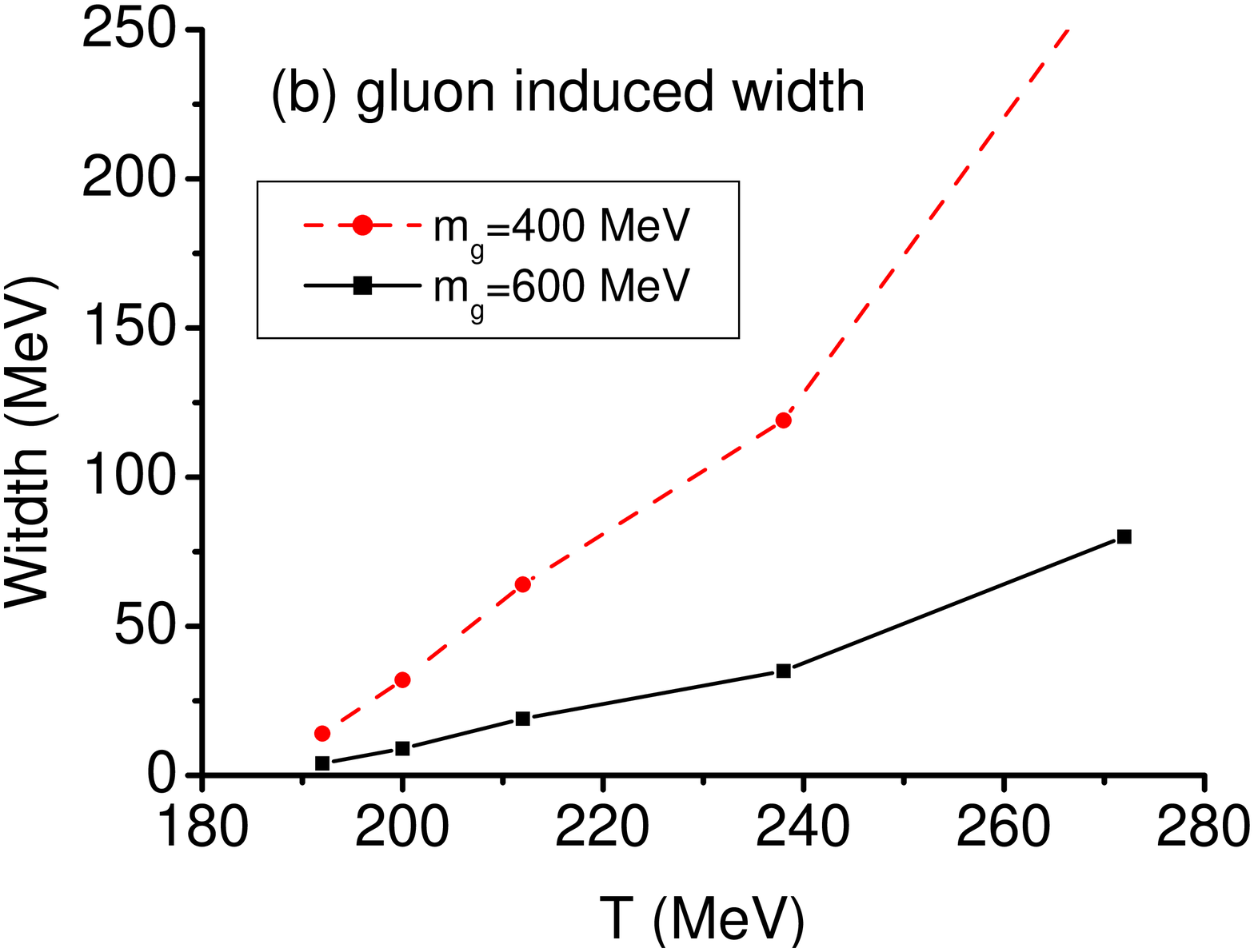}
\end{minipage}
\caption{\label{fig2} Effective thermal width due to (a) quarks
(b) gluons. Squares and circles are the results at temperatures
where the $J/\psi$ radius is calculated in ref\cite{Wong04}.}
\end{figure}


\section*{References}
\begin{thebibliography}{10}
\bibitem{Matsui86}
  Matsui T, Satz H 1986,
  Phys.\ Lett.\ B {\bf 178}, 416

\bibitem{Phenix1}
  Adare A, [PHENIX Collaboration] 2006,
  arXiv:nucl-ex/0611020

\bibitem{Gorenstein99}
  Gazdzicki M, Gorenstein M I 1999,
  Phys.\ Rev.\ Lett.\  {\bf 83}, 4009

\bibitem{PBM99}
  Braun-Munzinger P and Stachel J 2000,
  Phys.\ Lett.\ B {\bf 490}, 196

\bibitem{PBM06}
  Andronic A, Braun-Munzinger P, Redlich K and Stachel J 2006,
  arXiv:nucl-th/0611023

\bibitem{Hatsuda03}
  Asakawa M, Hatsuda T and Nakahara Y 2001,
  Prog.\ Part.\ Nucl.\ Phys.\  {\bf 46}, 459

\bibitem{Hatsuda04}
  Asakawa M and Hatsuda T 2004,
  Phys.\ Rev.\ Lett.\  {\bf 92}, 012001


\bibitem{Datta03}
  Datta S, Karsch F, Petreczky P and Wetzorke I 2004,
  Phys.\ Rev.\ D {\bf 69}, 094507

\bibitem{Datta05}
  Datta S, Karsch F, Petreczky P and Wetzorke I 2005,
  J.\ Phys.\ G {\bf 31}, S351

\bibitem{Lee87}
  Hansson T H, Lee S H and Zahed I 1988,
  Phys.\ Rev.\ D {\bf 37}, 2672

\bibitem{KS94}
Kharzeev D and Satz H 1994 {\it Phys. Lett.} B {\bf 334} 155;
Kharzeev D et al 1996 {\it Phys. Lett.} B {\bf 389} 595

\bibitem{AGGA01}
Arleo F, Gossiaux P-B, Gousset T and Aichelin J 2002 {\it Phys.
Rev.} D {\bf 65} 014005

\bibitem{Wong04}
  Wong C Y 2005,
  Phys.\ Rev.\ C {\bf 72}, 034906


\bibitem{Peskin79}
Peskin M E 1979 {\it Nucl. Phys.} {\bf B156} 365

\bibitem{BP79}
Bhanot G and Peskin M E 1979 {\it Nucl. Phys.} {\bf B156} 391

\bibitem{SL05}
  Song T and Lee S H 2005,
  Phys.\ Rev.\ D {\bf 72}, 034002

\bibitem{OKL02}
  Oh Y S, Kim S and Lee S H 2002,
  Phys.\ Rev.\ C {\bf 65}, 067901

\bibitem{LH97}
  Levai P and Heinz U W,
  Phys.\ Rev.\ C {\bf 57}, 1879 (1998)

\end{thebibliography}
\end{document}